\documentclass[11pt]{elsart}
\usepackage[dvips]{graphicx} 
\usepackage{latexsym} 
\usepackage{amssymb} 
\usepackage{amsmath}
\usepackage{amsfonts}

\begin{document}

\begin{frontmatter}

\title{Cryptanalyzing an improved security modulated
chaotic encryption scheme using ciphertext absolute value}

\author{G. \'{A}lvarez*},
\author{F. Montoya},
\author{M. Romera},
\author{G. Pastor}.

\address{Instituto de F\'{\i}sica Aplicada, Consejo Superior de
Investigaciones Cient\'{\i}ficas, Serrano 144, 28006 Madrid,
Spain}
\corauth[corr]{Telephone: +34-915 618 806; Telefax: +34-914
117 651; Email: gonzalo@iec.csic.es}

\begin{abstract}This paper describes the security weakness of a
recently proposed improved chaotic encryption method based on the
modulation of a signal generated by a chaotic system with an
appropriately chosen scalar signal. The aim of the improvement is
to avoid the breaking of chaotic encryption schemes by means of
the return map attack introduced by P\'{e}rez and Cerdeira. A
method of attack based on taking the absolute value of the
ciphertext is presented, that allows for the cancellation of the
modulation scalar signal and the determination of some system
parameters that play the role of system key. The proposed improved
method is shown to be compromised without any knowledge of the
chaotic system parameter values and even without knowing the
transmitter structure.

\end{abstract}


\end{frontmatter}

\bibliographystyle{unsrt}

\mathindent 0cm 
\sloppy 

\section{Introduction}
The possibility of synchronization of two coupled chaotic systems
was first shown by Pecora and Carrol
\cite{pecora90,pecora91,carroll91}. The importance of this
discovery was quickly appreciated \cite{he92,boccaletti02}, and
soon this topic aroused great interest as a potential means for
communications \cite{dedieu93,kocarev95,Tse03}. In recent years, a
considerable effort has been devoted to extend the chaotic
communication applications to the field of secure communications.
Accordingly, a number of cryptosystems based on chaos has been
proposed \cite{cuomo93a,cuomo93b,wu93,lozi93,yang04}; some of them
fundamentally flawed by a lack of robustness and security
\cite{short94,zhou97a,alvarez03,alvarez03b,alvarez04}.

P\'{e}rez and Cerdeira  showed in \cite{perez95} that it was
possible to retrieve the data encrypted by chaos when a parameter
of the sender is switched between two values. The data were
recovered directly from the ciphertext signal, without using the
authorized receiver and even without reconstructing the dynamics
of the chaotic system, by means of the attractor of a return map.
Later this method was further developed by Yang et al. in
\cite{yang98h}.

Recently Bu and Wang \cite{bu04} proposed an improved chaotic
encryption method. The aim of the improvement is to foil the
return map attack to chaotic encryption schemes of
\cite{perez95,yang98h}. The proposed approach is illustrated by
means of a Lorenz system, described by the following equations:
\begin{equation}\label{eq:Lorenz}
\rm \left\{ \begin{array}{l}
    \dot x_1 = \sigma(x_2-x_1) \\
    \dot x_2 = rx_1 - x_2 - x_1x_3 \\
    \dot x_3 = x_1x_2 - bx_3 \\
 \end{array} \right.
\end{equation}
being $\sigma$, $r$ and $b$ the system internal parameters.

In the literature, when digital encryption takes place, the signal
$s=x_1(t)$ is commonly taken as the transmitted encrypted message.
The authors of \cite{bu04} propose as ciphertext the modulation
signal $s(\textbf{x},t)=g(t)\,x_1(t)$.

The modulation signal $g(t)$ should be an appropriately chosen
scalar signal. For the Lorenz dynamical system they propose to
choose $g(t)$ as the product of a sinusoidal factor times a
variable of the chaotic system: $g(t)=A\cos(\omega t+\varphi_0)\,
x_3(t)$. Hence, the transmitted ciphertext will be:
$s(\textbf{x},t)=A\cos(\omega t+\varphi_0)\, x_1(t)\, x_3(t)$.

The key of the system is composed by the unknown values of the
system internal parameters together with the, so called, external
parameters $\omega$ and $\varphi_0$, thus considerably enlarging
the key space with respect to the simple chaotic encryption.

The authors of \cite{bu04} show that the attractor of the return
map of the ciphertext is blurred by the modulation and,
consequently, they claim that an intruder can not retrieve any
information from it. They also show that the Fourier power
spectrum of the ciphertext does not allow for the knowledge of the
frequency $\omega$ of the sinusoidal factor.

The authors seem to base the security system on the impossibility
of performing a return map attack and determining the sinusoidal
modulating signal; but no general analysis of security is
included. The weaknesses of this system and a method to break it
are discussed in the next section.

\section{Ciphertext absolute value attack}

In this section is presented an attack to the proposed improved
security encryption scheme based on taking the absolute value of
the ciphertext. Our attack allows for the system breaking in two
different ways. The first one consists of retrieving the plaintext
directly from the ciphertext, when digital encryption is used. The
second one consists of the identification and cancelation of the
modulating signal $\cos(\omega t + \varphi_0)$, allowing the
plaintext recovery by the return map attack, thus circumventing
the alleged system security improvement.

The Lorenz attractor has a complicated shape, with trajectories
spiralling around, and jumping between two loops, either in three
dimensions or in any two-dimension projection.

\begin{figure}[t]
  \center
  \includegraphics{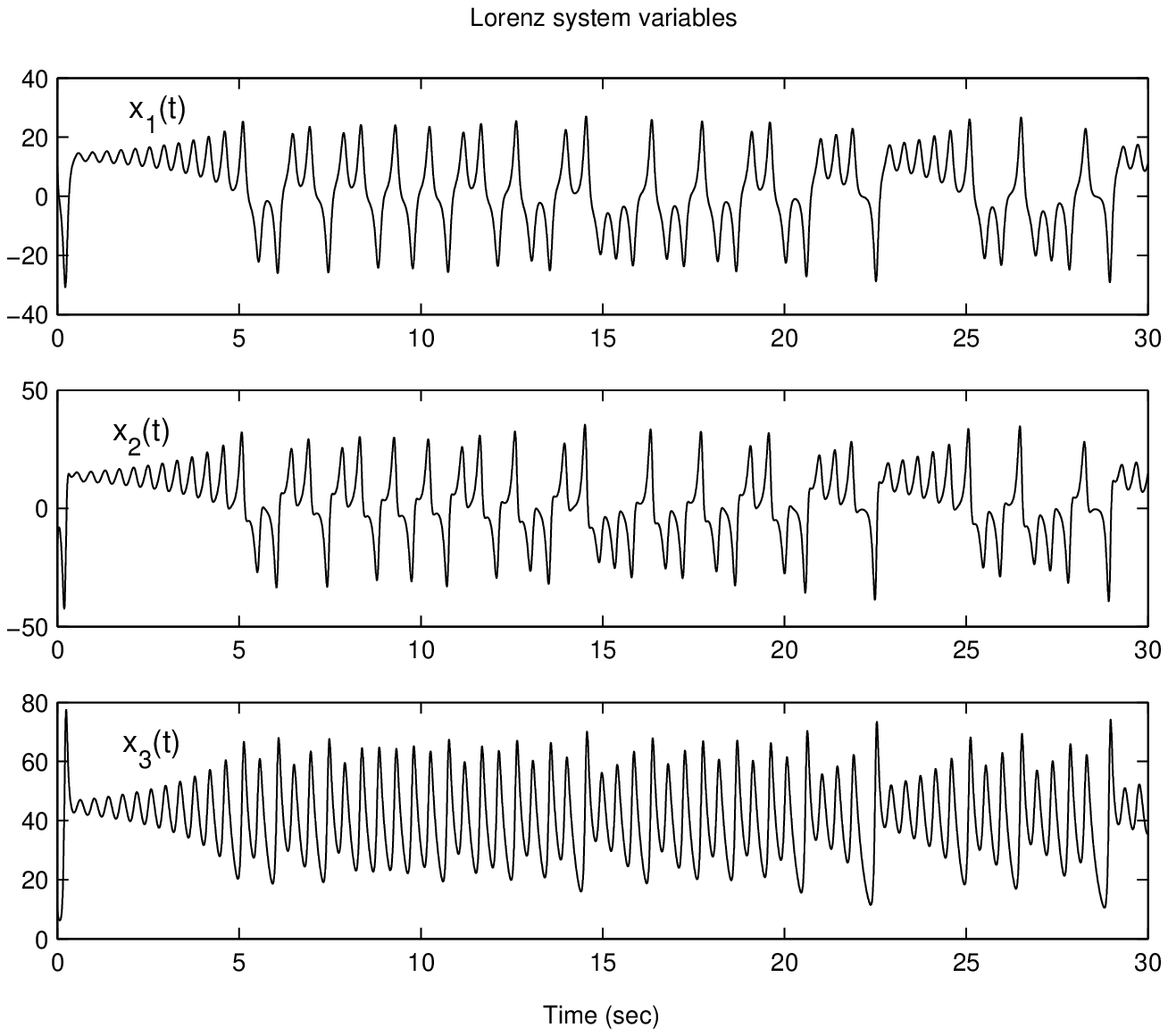}
  \caption{Time story of the variables $x_1(t)$,
  $x_2(t)$, and $x_3(t)$ of of the Lorenz system.}
  \label{fig:timestory}
\end{figure}

It is interesting to look at the time story and frequency power
spectrum of each component of the Lorenz system. As illustrated in
Fig.~\ref{fig:timestory}, the variable $x_3(t)$ shows a quasi
sinusoidal oscillation, always of positive value, whose amplitude
changes in an irregular sawtooth like fashion; while the variables
$x_1(t)$ and $x_2(t)$ show a similar behavior but with sudden
large amplitude jumps across the zero axis, corresponding to the
attractor jumps between loops.

\begin{figure}[t]
  \center
  \includegraphics{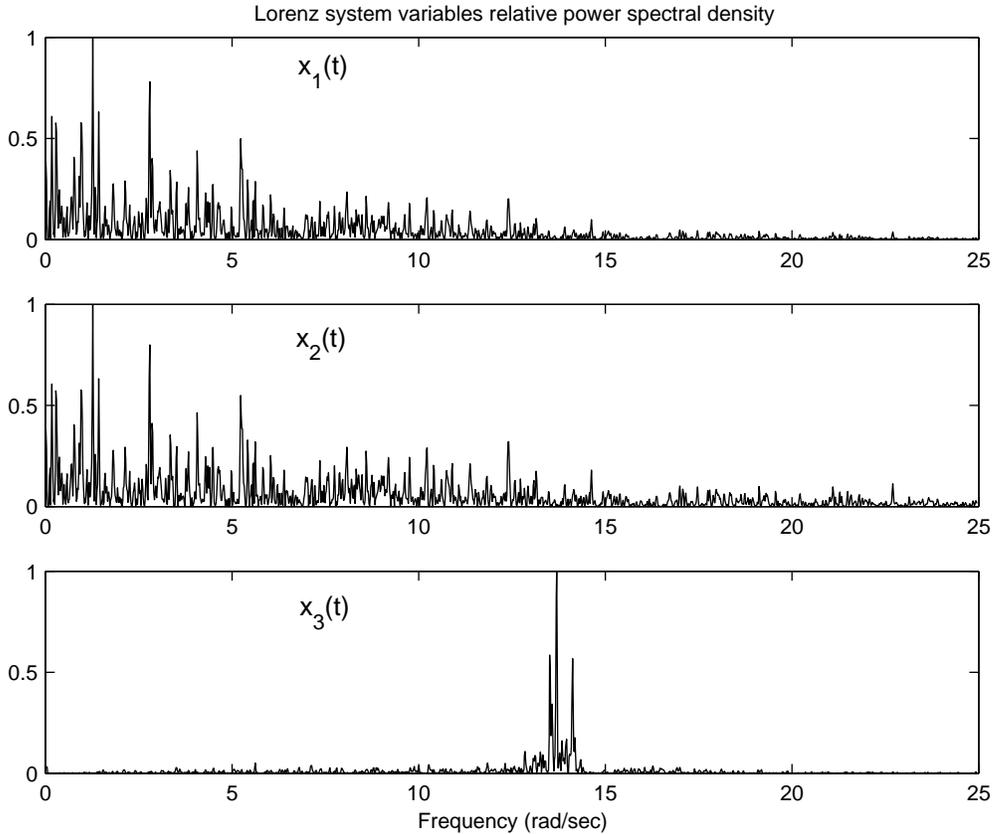}
  \caption{Power spectral density of the variables $x_1(t)$,
  $x_2(t)$ and $x_3(t)$ of the Lorenz system.}
  \label{fig:3spectra}
\end{figure}

The frequency power spectra of the Lorenz system variables are
illustrated in Fig.~\ref{fig:3spectra}. As observed, $x_3(t)$ has
a relative clean and simple spectrum, with a narrow band at 13.7
rad/sec, that corresponds to the stable frequency of its
oscillatory waveform. However, the variables $x_1(t)$ and $x_2(t)$
have a much broader and complex spectra, with maximum amplitude
near 3 rad/sec, due to the aperiodic jumps across the zero axis.

\begin{figure}[h]
  \center
  \includegraphics{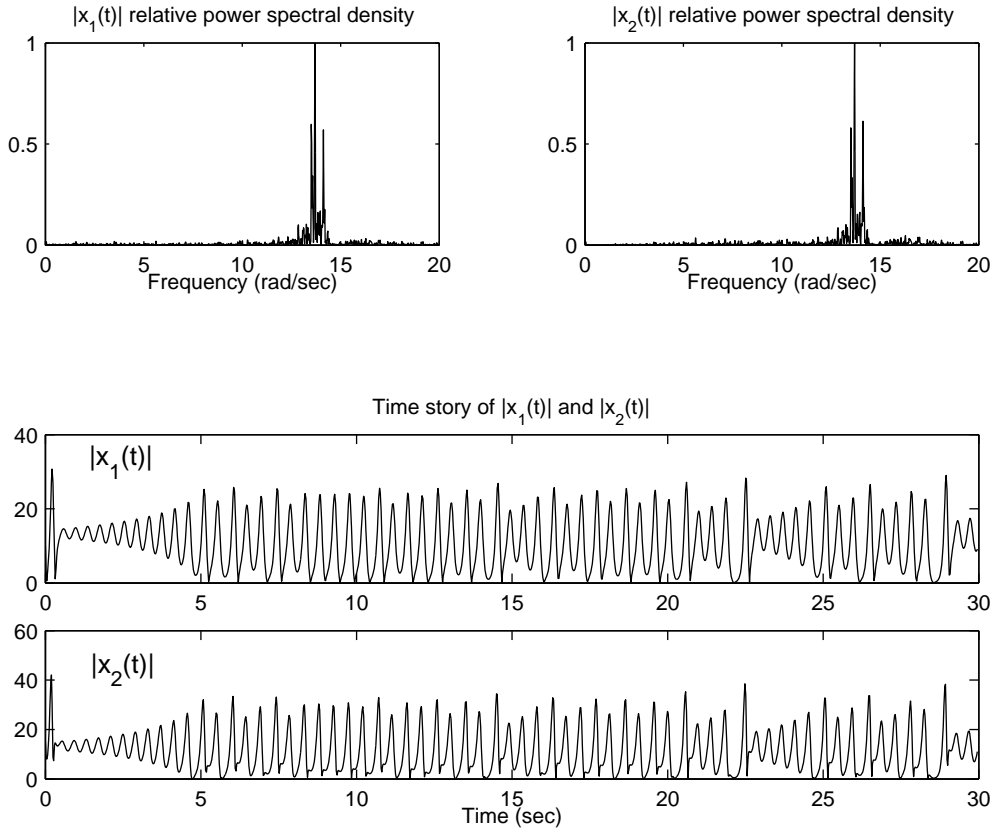}
  \caption{Power spectral density and time story of $|x_1|$ and
  $|x_2|$ of the Lorenz system.}
  \label{fig:4absolut}
\end{figure}

Let us take the absolute value of the two first Lorenz system
variables: $y_1(t)=|x_1(t)|$, $y_2(t)=|x_2(t)|$. When computing
the time stories and frequency power spectra of $y_1(t)$ and
$y_2(t)$ the results illustrated in Fig.~\ref{fig:4absolut} are
obtained. Both time stories resemble very much that of $x_3(t)$,
while their frequency power spectra are practically
indistinguishable from that of $x_3(t)$. That is, a new set of
signals $y_1(t)$ and $y_2(t)$ has been built, whose waveform and
spectra are much simpler than the ones of the original variables.
In fact, when substituting $x_1(t)$ or $x_2(t)$ with $y_1(t)$ or
$y_2(t)$ the double scroll is converted into a simple scroll with
quite regular motion.

It is assumed in the literature that chaotic modulation is an
adequate means for secure transmission, because chaotic maps
present some properties as sensitive dependence on parameters and
initial conditions, ergodicity, mixing, and dense sinusoidal
points. These properties make them similar to pseudorandom noise
\cite{devaney92}, which has been used traditionally as a masking
signal for cryptographic purposes.

A fundamental requirement of the pseudorandom noise used in
cryptography is that its spectrum should be infinitely broad,
flat, and of higher power density than the signal to be concealed
within. In other words, any information power spectrum should be
buried into the pseudorandom noise power spectrum.

If any of the Lorenz system variables $x_1(t)$ or $x_2(t)$ is used
as a carrier signal for secure transmission, an attacker can take
its absolute value to build a new signal with much simpler shape
and spectrum. This new signal does not satisfy the above
conditions. As a consequence, it may still preserve some plaintext
information within, certainly alleviating the task of breaking the
system.

\subsection{Direct plaintext recovery}
\label{subsec:direct} The main problem with this kind of
cryptosystems lies on the fact that the ciphertext is an analog
signal, whose waveform depends on the system parameter values.
When a parameter is switched between two different values by a
digital plaintext \cite{Parlitz92b}, the system changes between
two different attractors with different waveforms, amplitudes and
frequencies. Therefore it is possible to extract some information
by a suitable signal processing of the ciphertext.

\begin{figure}[t]
  \center
  \includegraphics{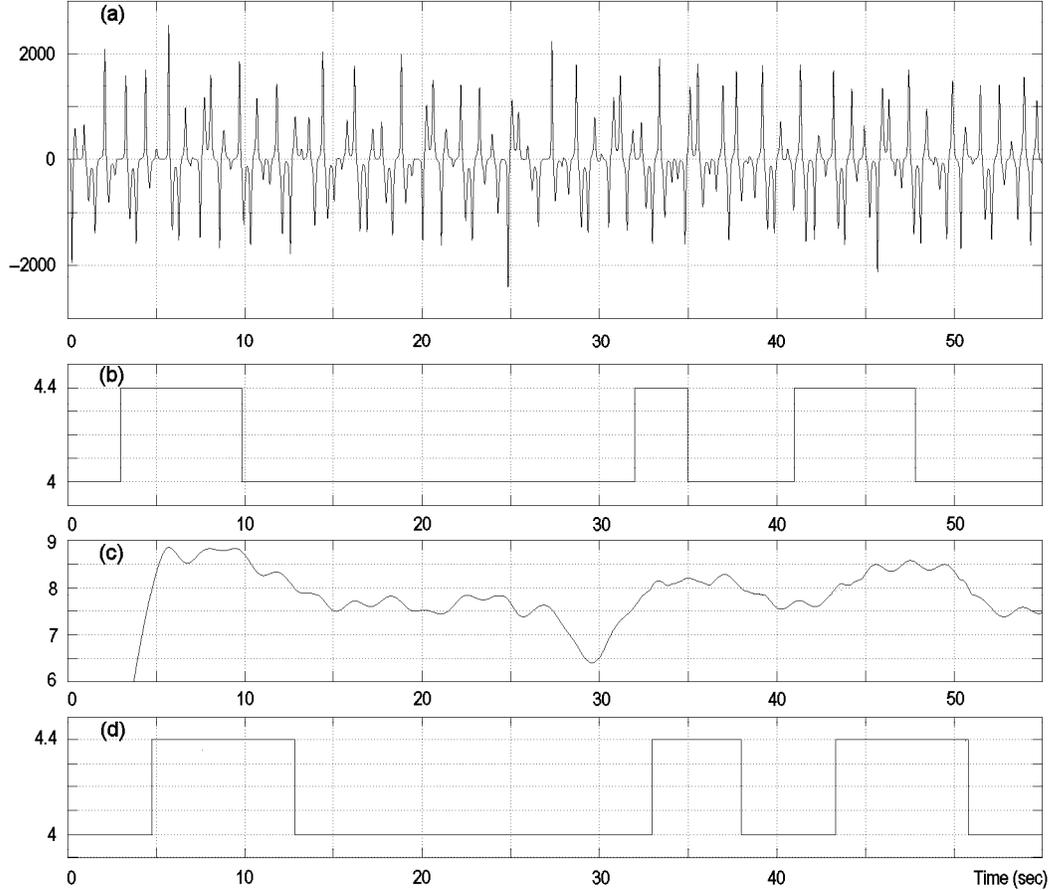}
  \caption{Full wave amplitude detector: (a)ciphertext signal;
  (b) plaintext; (c) low-pass filtered absolute value of the
ciphertext signal; (d) recovered plaintext.}
  \label{fig:direct}
\end{figure}

To illustrate the possibility of breaking the system, the chaotic
transmitter of \cite[\S 2~and~fig. 4]{bu04} has been simulated
with the same parameter values of the example, that is: $(A,
\omega, \varphi_0) =(0.5, 1.5, 0.0)$, being the parameter $b$
changed between its reference value $b_0=4.0$ when the values 0's
are to be transmitted and $b_1=4.4$ when the values 1's are to be
transmitted. The parameter values have been arbitrarily chosen as:
$(\sigma, r)= (16, 45.6)$ and the system initial conditions  as:
$(x_1(0), x_2(0), x_3(0))=(-5.8, 2, 2.41)$. Our simulation was
done with a four order Runge-Kutta integration algorithm in
MATLAB~6 with a step size of $0.001$.

To break the system a so called full wave amplitude detector has
been implemented, consisting of taking the absolute value of the
ciphertext signal $s(\textbf{x},t)=A\cos(\omega t+\varphi_0)
\,x_1(t)\, x_3(t)$. Next, this signal is low-pass filtered and,
finally, binary quantized. The low-pass filter employed is a six
pole Butterworth with a frequency cutoff of 1 rad/sec. The
quantizer is a level comparator with switch point at level value 8
of the low-pass filtered absolute value of the ciphertext signal.

The procedure is illustrated in Fig.~\ref{fig:direct}. The result
is a good estimation of the plaintext, with tiny inaccuracies
consisting of small delays in the transitions, due to the time
delay introduced by the filter. It should be emphasized that our
analysis is a blind detection, made without the least knowledge of
what kind of non-linear time-varying system was used for
encryption, nor its parameters values and neither its keys.

\begin{figure}[t]
  \center
  \includegraphics{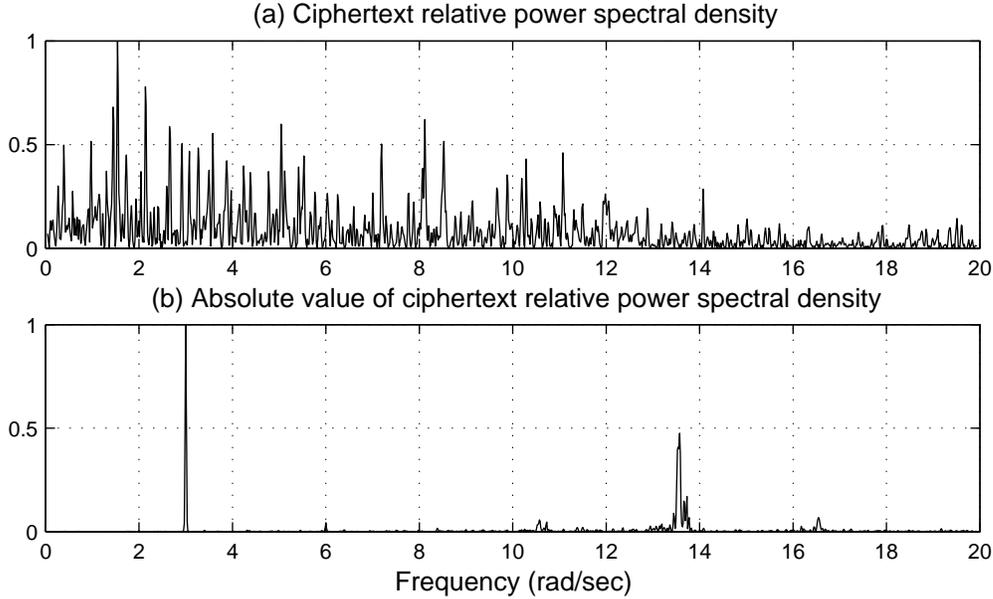}
  \caption{Power spectral density of the ciphertext signal:
  (a) spectrum of the transmitted signal
  $A \cos(\omega t+\varphi_0)\, x_1(t)\,x_3(t)$;
  (b) spectrum of the absolute value of the transmitted signal
  $|A \cos(\omega t+\varphi_0)\, x_1(t)\,x_3(t)|$.}
  \label{fig:cipherspectr}
\end{figure}

\subsection{Modulating signal cancellation}

In the encryption method proposed in \cite{bu04} the ciphertext
signal $s(\textbf{x},t)=A \cos(\omega t+\varphi_0)\, x_1(t)\,
x_3(t)$ shows a crowded frequency power spectrum as is illustrated
in Fig.~\ref{fig:cipherspectr}(a). As the authors  point out, it
is impossible to identify in it any separated factor that can be
exploited to attack the system. This is a consequence of the use
of the Lorenz system and the modulating procedure. First, the
spectrum of $x_1(t)\, x_3(t)$ is a complicated one, due to the
chaotic nature of the Lorenz system. Second, the multiplication of
this signal by $A\cos(\omega t+\varphi_0)$ has the effect of
generating a new spectrum composed by the superposition of two
versions of the $x_1(t)\, x_3(t)$ spectrum shifted $\pm\,\omega$
radians per second. And third, the value chosen for $\omega$ falls
in the neighborhood of the maximum of the spectrum of $x_1(t)\,
x_3(t)$, hence it is impossible to identify it from the frequency
power spectrum.

\begin{figure}[t]
  \hspace{-1.2cm}
  \includegraphics{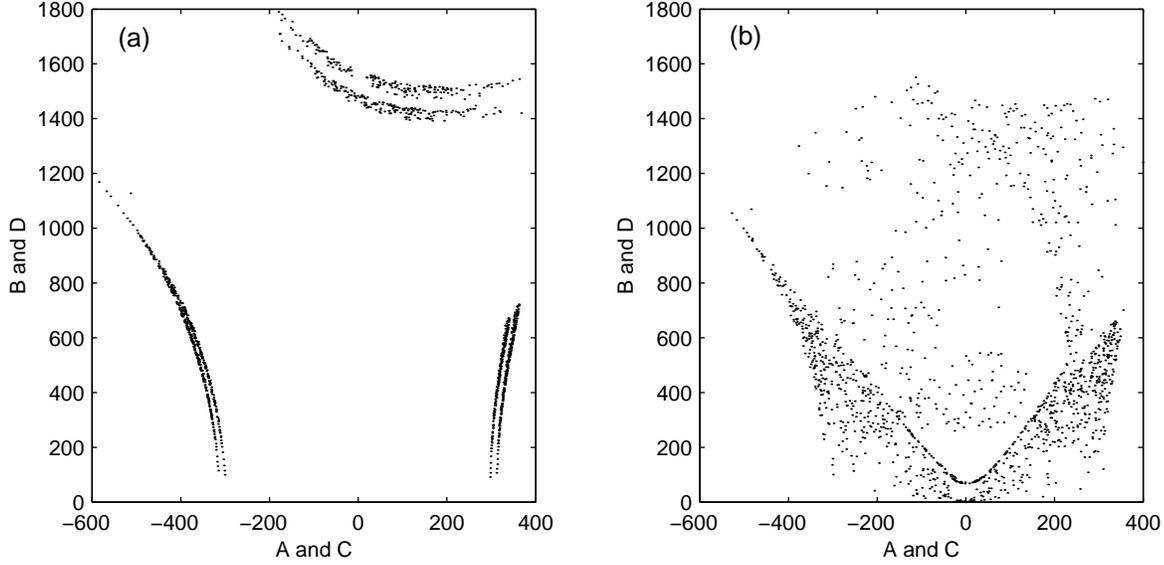}
  \caption{Attractor of the return map of the ciphertext signal: (a) clean attractor of an ideal signal
  with no sinusoidal factor $0.5\, x_1(t)\,x_3(t)$;
  (b) blurred attractor of the actual transmitted signal with
  sinusoidal factor
  $0.5 \cos (1.5\,t)\, x_1(t)\,x_3(t)$.}
  \label{fig:return}
\end{figure}

The aim of this encryption method is to foil the return map
attack, by blurring the attractor of the return map. In
Fig.~\ref{fig:return} the attractor of the transmitted signal
$s(\textbf{x},t)$ is plotted when digital encryption takes place
in two different cases. In Fig.~\ref{fig:return}(a) without
sinusoidal factor:
\begin{equation}
s(\textbf{x},t)=x_1(t)\, x_3(t),
\end{equation}
and in Fig.~\ref{fig:return}(b) when the sinusoidal factor is
present:
\begin{equation}
s(\textbf{x},t)=A \cos(\omega t+\varphi_0)\, x_1(t)\,x_3(t),
\end{equation}
where $(A,\omega, \varphi_0) =(0.5, 1.5, 0.0)$, being the system
parameters for encryption  $(\sigma, r, b_0, b_1)= (16, 45.6, 4.0,
4.4)$. It can be seen that when there is no sinusoidal modulation
factor it is possible to use the return map attractors to retrieve
the plaintext, because it is clearly distinguishable the splitting
of the attractor segments originated by the plaintext presence.
But when there is a sinusoidal modulating factor the attractors
happen to by fully blurred, obstructing any straightforward
analysis.

To break such encryption scheme the following procedure is
proposed, which will allow for the reconstruction of the return
map attractor.

First the frequency of the sinusoidal factor is identified from
the ciphertext by computing the Discrete Fourier Transform of the
ciphertext absolute value. Let $\omega'$ be the approximate value
of $\omega$ and $\varphi'_0$ the approximate value of $\varphi_0$.
As is illustrated in Fig.~\ref{fig:cipherspectr}(b) it can be seen
that there is a very neat pick in the spectrum exactly at
$2\omega'=3$ rad/sec, which corresponds to the double of the
sinusoidal factor frequency. This is due to the non linear nature
of the absolute value function, which has the effect of folding
the ciphertext upwards around the zero axis, effectively doubling
the frequency of the sinusoidal factor. The part of the spectrum
corresponding to the Lorenz system factor,
$s(\textbf{x},t)=A\,x_1(t)\, x_3(t)$, appears as a band of
frequencies near 13.7 rad/sec. Finally, there are two small
amplitude side bands at $13.7\pm3$ rad/sec, corresponding to the
cross-modulation between the sinusoidal modulation factor and the
Lorenz system factor. The spectrum was calculated using a
16384~point Discrete Fourier Transform with a 4-term
Blackman-Harris window.

\begin{figure}[t]
  \includegraphics{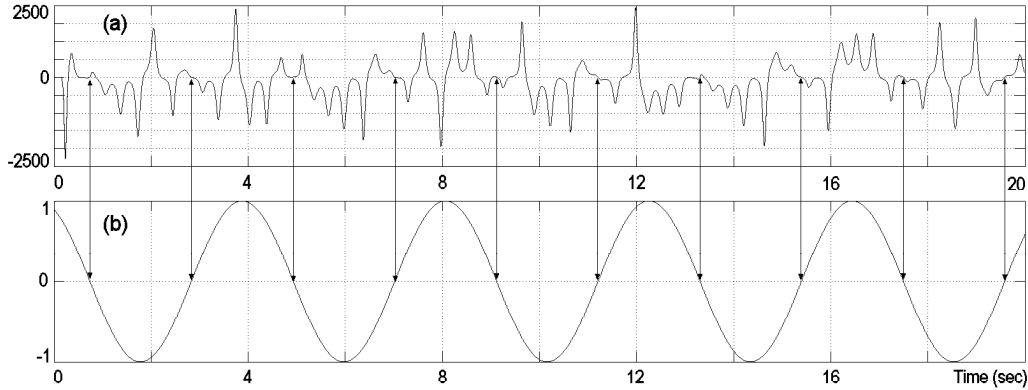}
  \caption{Zero crossing phase identification of $\varphi'_0$}
  \label{fig:ceros}
\end{figure}

Once the sinusoidal factor frequency $\omega'$ is estimated, it is
a trivial task to determine its phase $\varphi'_0$ just searching
for the ciphertext zero crossing points at intervals equal to
$\pi/4\omega'$, as depicted in Fig.~\ref{fig:ceros}.

Next, the approximate cancellation of the sinusoidal factor must
be performed.  Let $x'_1$ and $x'_2$ be the approximate values of
$x_1$ and $x_2$. Should the determination of $\omega'$ and
$\varphi'_0$ be exact it would be satisfactory to divide the
ciphertext signal s(\textbf{x},t) by the estimated sinusoidal
factor $cos(\omega't+\varphi'_0)$ to get rid of it, as:
\begin{equation}\label{eq:simple}
 A\, x'_1(t) \,x'_3(t)=\frac{A\cos(\omega t+\varphi_0) x_1(t) x_3(t)}
 {\cos(\omega' t+\varphi'_0)}
\end{equation}
But, in practice, small inaccuracies in the determination of
$\varphi'_0$ can lead to a division by cero, with infinite
amplitude error. If the error $\varepsilon=\varphi'_0-\varphi_0$
is $\varepsilon\geq0.001$ radians it would be better to add some
constant value $\eta$ around the zero level of the estimated
sinusoidal factor, as follows:
\begin{equation}\label{eq:exact}
 A\, x'_1(t)\, x'_3(t)=\frac{A\cos(\omega t+\varphi_0) x_1(t)
 x_3(t)} {\mathrm{sgn}[\cos(\omega' t+\varphi'_0)]\, |\cos(\omega' t+
 \varphi'_0) +\eta|}
\end{equation}
\begin{figure}[t]
  \hspace{-1 cm}
  \includegraphics{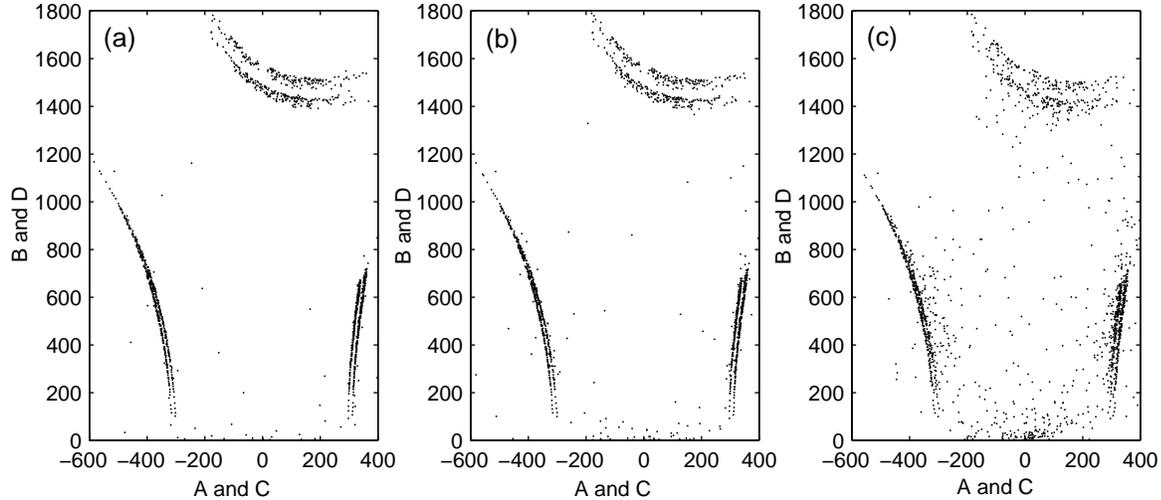}
  \caption{Reconstruction of the attractor of the
  return map of the ciphertext signal digitally modulated,
  for various values of the error
  $\varepsilon=\varphi'_0-\varphi_0$: (a) $\varepsilon=0,0005$
  radians; (b) $\varepsilon=0.001$ radians; (c)
  $\varepsilon=0.01$ radians;}
  \label{fig:recuperation}
\end{figure}
Finally, once the cancellation of the sinusoidal modulation has
been performed, the return map attractor can by plotted. After
simulating the reconstruction of the return map attractor of the
example of \ref{subsec:direct}, the results are presented in
Fig.~\ref{fig:recuperation}. It can be seen that the return map
attractor is reconstructed with enough accuracy to enable the
return map attack described in \cite{perez95,yang98h}, thus
deceiving the improved security claimed by the authors
of~\cite{bu04}. In our simulation Eq.~(\ref{eq:simple}) was used
to reconstruct the attractor with a $\varphi_0$ determination
error value of $\varepsilon=0,0005$ radians; while for errors
values of $\varepsilon=( 0.001, 0.01)$ radians, equivalent to 0.06
and 0.6 degree, Eq.~(\ref{eq:exact}) was used, with $\eta$ values
of 0.001 and 0.005, respectively.

\section{Conclusion}
The lack of security of the encryption method proposed in
\cite{bu04} is made evident, since it can be broken without
knowing its parameter values and even without knowing the
transmitter's precise structure. A new method that can retrieve
the plaintext and determine part of the key directly from the
ciphertext has been devised. As a result of our attack, the
security of the proposed encryption scheme does not improve over
the traditional chaotic encryption methods, being still vulnerable
to the same attacks it intends to foil.

\ack{This work was supported by Ministerio de Ciencia y
Tecnolog\'{\i}a of Spain, research grants TIC2001-0586 and
SEG2004-02418}.

\end{document}